\documentclass[12pt,fleqn]{article}
\usepackage{graphicx}
\begin{document}
\title{\bf A Few Exactly Solvable Models For Riccati Type Equations}
\author{\bf B G Sidharth and B S Lakshmi\\Center For Applicable Mathematics and
Computer Sciences,\\B M Birla Science Center,\\\small Adarsh
Nagar, Hyderabad-500063, India}
\date{}
\maketitle
\begin{abstract}
We consider the Ricatti equation in the context of population
dynamics,quantum scattering and a more general context.We examine
some exactly solvable cases of real life interest.
\end{abstract}
\section{\bf Introduction}
The Riccati Equation \cite{weig,hart,ross}
\begin{equation}
y' = p(t) + q(t) y + r(t) y^2 \end{equation} appears in several
branches of applicable mathematics, for example population
dynamics and mathematical physics, as in quantum scattering.It is
known that it exhibits chaotic behavior \cite{cook}.It is also well known that,
it reduces to a second order  linear differential equation by the substitution,
 $y = \frac{u'}{u} $,\\
when we get, \begin{equation} u'' + g(t)u' + f(t) u = 0
\end{equation} (1) and (2) have been studied in detail for a long
time\cite{reid,sid,lak}.\\It may be mentioned that if two solutions of the
Riccati equation are equal at a point, then they coincide.The
reason lies in the continuity of the logarithmic derivative
$\frac{u'}{u}$ given in the substitution that lead from (1) to
(2).\\We will now consider exact solutions of the Riccati equation
in two different contexts,from the field of population dynamics
and the field of quantum scattering.
\input epsf
\section{Some exact solutions}
\subsection{A Problem In Population Dynamics}
We now examine coupled Riccati type equations of relevance to a population
model.
\subsubsection*{Gause's Model}
Consider two-species populations occurring together,and assume
that the growth of each is inhibited by members ,both of its own
and of the other species.Denoting the number of individuals in
species \rm{1}  as $N_1$ and species \rm{2} as $N_2$,we have the Gause's
competition equations:\cite{pie}
\begin{equation}\frac{1}{N_1}
\frac{dN_1}{dt} = r_1- a_1 N_1 - a_2 N_2\end{equation}
\begin{equation}\frac{1}{N_2} \frac{dN_2}{dt} = r_2-a_3 N_1 - a_4 N_2
\end{equation}
where $r_{1}$,$r_{2}$,$a_{1}$,$a_{2}$ are defined below. Thus we
are assuming that the per capita growth of each population at an
instant is a linear function of the sizes of the two competing
populations at that instant.Each population would grow
logistically if it were alone with logistic parameters $r_1$ and
$a_{1}$ for species \rm{1} and $r_{2}$ and $a_{2}$ for species
\rm{2} .\\ In general the simultaneous differential equations
cannot be explicitly solved.We now consider a particular set of
circumstances in which they can be solved.Thus for (3) and (4) we
specialize to, with a more convenient notation ,
\begin{equation}
\frac{dy}{dt}= 1-xy^{2}
\end{equation}
\begin{equation}
\frac{dx}{dt}= 1+yx^{2}
\end{equation}
From (5) and (6) we get,
\begin{equation}
\frac{dy}{dx}= \frac{1-xy^{2}}{1+yx^{2}}
\end{equation}
Integrating (7) we have
\begin{equation}
y-x = -x^{2}y^{2}+ C
\end{equation}
From (8) when $y = 0$,$C=-x_{0}$ so that  $y < x$ \\Reverting back
to $N_{1}$ and $N_{2}$, this is \begin{equation} N_{1}-N_{2} =
-{N_{2}}^2{N_{1}}^2+ C \end{equation} From (9) when
$N_{1}=0$,$C=-N_{0}$ so that $N_{1} < N_{2}$.\\The Figure below
illustrates a particular case of the
above solution with the two populations $N_{1}$ and $N_{2}$ along Y and X axes.\\

\includegraphics[bb= 0 0 224 138]{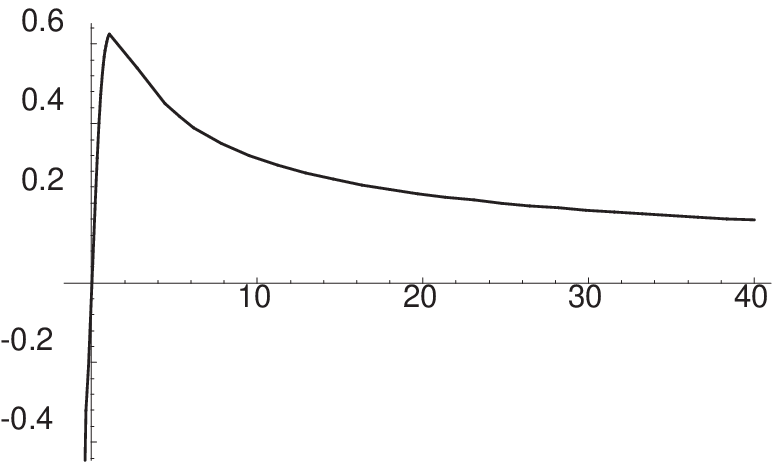}

\subsection{Problems from Scattering Theory}
By reversing the method used above ,the quantum mechanical radial
Schr\"{o}dinger equation
\begin{equation}
u'' - H(r) u = 0
\end{equation}
 can be reduced to a Riccati equation
\begin{equation}
  v' + v^{2}= H(r)
\end{equation}
by the substitution $v = \frac{u'}{u}$. The form (10)or (11) is
used in phase shift analysis for example in Calogero's variable
phase approach \cite{sidh}.It is also possible to use the form
(11) for building up an iterative procedure\cite{cal} .We would
now like to point out that (11) can be used in  a different
context for providing exact solutions for specified classes of the
potential function f(r). Let us write in \nolinebreak (11)  \[v
\equiv \frac{1}{f}\] \[ f'=1-gf
\]   Whence we get
\begin{equation}
  H= \frac{g}{f}=\frac{ge^{\int{g}dr}}{\int{e^{\int{g}dr}+C}}
\end{equation}
1.We put g=$\alpha$ in (12) \[ H=\frac{\alpha^{2}}
{1+c\alpha e^{-\alpha r}}\]
The potential H is now of
the form \[
 H =\frac{D}{A+Be^{\alpha r}}
 \]
This is the well-known Wood-Saxon potential. The solution u is
given by
\[
  u=C_{1}(e^{\alpha r}+C_{2}\alpha)
\]
 2.Putting g=r
in (12), the potential H is given by
\[ H= \frac{r e^{r^2/2}}{\int e^{r^2/2} dr +
c_{1}}\]
This is a modified Gaussian potential and the solution u is given by
\[
  u = K \int e^{-r^2}dr + C_{3}
\]
 The Graph is shown in the figure
below and is seen to fall very steeply indicative of a confined state or
particle.\\ \vspace{9 pt} \\
\includegraphics[bb= 0 0 224 138]{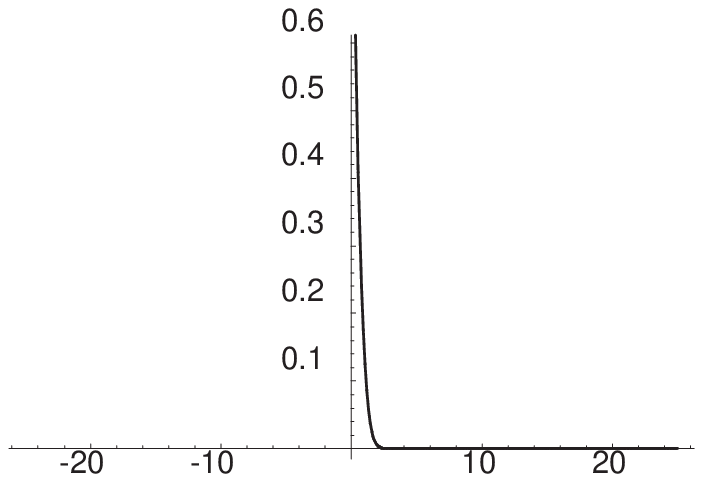}

 3.Putting g=1/r in
(11),the potential H is given by
\[ H= \frac{2}{r^2+c_{4}}
\]
This is a shifted inverse square potential and the
solution u is given by
\[
  u= K'(r^2 +C_{4})
\]
\section{Other Exact and Asymptotic Solutions}
We now consider some asymptotic solutions of (11),which we write as,
\begin{equation}
u' = u^{2} + f \end{equation}
 Let \[ \int{f} dr \equiv {g}, \]
f and g being bounded functions.
So (13) can be written as ,
\begin{equation} z' = (z+g)^2 > 0, z=u-g, \end{equation}
which shows that z is an increasing function of r.Suppose z is
unbounded.So for large r,we should have from (14)
\[ z \approx z^2 \]
whence \[ z = \frac{1}{c-r} \] which $ \rightarrow 0 $ as $ r
\rightarrow \infty $.\\ This is a contradiction. Therefore z is
bounded and so also u ,that is $ u\longrightarrow M
$\quad as \quad $r \longrightarrow \infty $\\
Therefore $u'\approx 0 $ (asymptotically)\\
Therefore for large r,(13) becomes
 \[ u^{2} = -f, \textup{whence} \]
\begin{equation}  u = \pm (-f)^{1/2} \end{equation}
By way of verification of (15),let us consider (13) with,
\[ f = - \frac{\alpha^2}{r^{2}}\]
So we expect that for large r, $u \sim$ $ \pm$ $ \frac{\alpha}{r}$
by (15). Let us put  \begin{equation} u = \frac{\beta}{r}
\end{equation} in (13). So \[-\frac{\beta}{r^{2}} =
\frac{\beta^{2}}{r^{2}} -\frac{\alpha{2}}{r^{2}} \] or \[ \beta =
\frac{-1 \pm\sqrt{1+4\alpha^{2}}}{2} \approx \pm
\alpha\hspace{3mm} \textup{if} \hspace{3mm}\alpha \ \gg 1 \]
  \textup{That is} \[ u = \pm
\frac{\alpha}{r} \] (everywhere, and so  also for large r). Using
this example,with transformations of the independent variable,we
can generate similar solutions. For example, if we substitute for
r, $ t = t(r)$, (13) becomes
 \[ t'\dot{u} = u^2 + f,\]
where the dot denotes the derivative with respect to t.The choice
$t'=f(r)$, leads to a similar equation, and one can verify that
for
\[f(r) = \frac{a}{3} r^{-2/3} - a^2 r^{2/3},\hspace{5mm}  u = a
r^{1/3}\] is a solution.More generally as can be easily verified
$a r^n$ is a solution for $ f(r)=n a r ^{n-1}-a^2 r^{2n}$ and so
on.However in these examples,
neither u nor u' are asymptotically bounded.\\
Finally it maybe observed that if one solution of the Riccati equation
 (13) is known,then others could be derived therefrom\cite{ross}.
 
\end{document}